\documentclass[12pt,letterpaper]{report}

\usepackage{graphicx}
\usepackage[margin=1in]{geometry}
\usepackage{amsmath}
\usepackage[T1]{fontenc}
\usepackage[utf8]{inputenc}
\usepackage{authblk}
\usepackage{fancyhdr}
\usepackage{lastpage}
\usepackage[parfill]{parskip}
\usepackage{subcaption}
\usepackage{url}
\usepackage{apacite}

\begin{document}

\title{Interactive Exploration of the Employment Situation Report: From Fixed Tables to Dynamic Discovery}

\author[1]{Peter Mancini}
\author[2]{Benjamin Bengfort}
\author[2]{Ben Shneiderman}
\affil[ ]{University of Maryland}
\affil[ ]{Department of Computer Science}
\affil[ ]{A.V. Williams Building}
\affil[ ]{College Park, MD 20742}
\affil[ ]{United States of America}
\affil[ ]{}
\affil[1]{\textit{pmancini@umd.edu} -- corresponding author}
\affil[2]{\textit{\{bengfort,ben\}@cs.umd.edu}}

\date{\today}

\maketitle

\begin{abstract}
The monthly Bureau of Labor Statistics \textit{Employment Situation Report} is widely anticipated by economists, journalists, and politicians as it is used to forecast the economic condition of the United States. The report has broad impact on public and corporate economic confidence; however, the online access to this data employs outdated techniques, using a PDF format containing solely text and fixed tabular information. Creating an interactive interface for dynamic discovery on the BLS website could elicit more dialogue between the public and government spheres, drawing more traffic to government websites and triggering greater civic engagement. Our work suggests that the implementation of interactive visual analysis techniques to enable dynamic discovery leads to rapid interpretation of data as well as provides the means to explore the data for further insights. This paper presents two inspirational prototypes: a dashboard of interactive visualizations and an interactive time series explorer, allowing for temporal and spatial analyses and enabling users to combine data sets to create their own customized visualization. 

\end{abstract}

\section{Introduction}

Since 1984, the Bureau of Labor Statistics (BLS) has served as the principal Federal agency responsible for measuring labor market activity, working conditions, and price changes in the economy. Data collected by the BLS can be sorted into four categories: Prices (e.g. Consumer Price Index), employment and unemployment, compensation and working conditions, and productivity. This data is collected via surveys designed by economists and statisticians to acquire information regarding demographic and industrial statistics, which are then released in a report to the public.

Results of the surveys are compiled in the \textit{Employment Situation Report}. This critical, monthly assessment of the state of labor in the United States is released by the Bureau of Labor Statistics (BLS) at 8:30 AM on the first Friday of the month. The ``Jobs Report'', as it is popularly called, is based on the Current Population Survey, which surveys individual households and the Current Employment Statistics Survey, which surveys employers. Together, these two surveys give a snapshot of the number of employed and unemployed Americans, how many hours they are working, and a variety of other facts and figures. Recently, for example, data from this report was used to measure economic and political solutions to the 2008 downturn, as well as the performance of politicians themselves during that time. As a result of such insights, news media widely publicizes the upcoming release of the report each month and follows the publication with widespread analyses and critiques. The report is more recently being used as a forecasting tool by firms on Wall Street, largely because the report is an indicator of investor and employer confidence in the economy \cite{mahorney_what_2013}.

The report is delivered on the BLS website in both PDF and HTML format, comprised mostly of text and tabular information and is well suited for deep study rather than rapid analysis. Because of the report's forecasting power and journalistic impact, a holistic interpretation of the report's information is essential; however, the current format does not lend itself well to the quick derivation of insights. Moreover, the report is not as easily accessible to a general audience, favoring economic language and measurements. A visual analytics methodology \cite{keim_mastering_2010} using information visualization techniques could benefit the \textit{Employment Situation Report}, allowing for swift interpretation of the current employment situation in the United States.

\begin{figure}[t]
	\centering
    \includegraphics[width=0.9\textwidth]{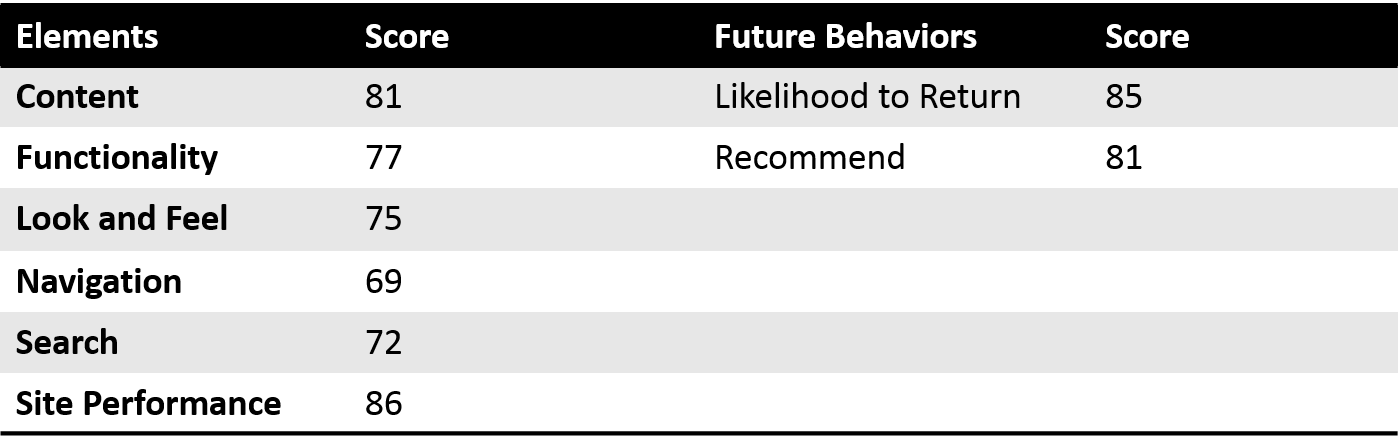}
    \caption{BLS customer satisfaction survey for July, 1 2009 - June, 30 2011.}
    \label{CustSat}
\end{figure}

In order to assess the comprehensibility and utility of their website, the BLS, characteristically, conducts a survey on customer satisfaction. Figure \ref{CustSat} provides the customer satisfaction results from July 1, 2009 - June, 30 2011. The average satisfaction level was 75, which implies that the website has  room for improvement. Our work aims to increase customer satisfaction and provide a more satisfying  and dynamic experience for the data-seeking users. We gather data contained in thousands of tables and present it so that is easily navigable and enables users conduct visual analytic exploration of the dataset.

The tools presented here were designed to be easy to use, visually clear, and capable of evoking insights by users. It is the designer's job to manipulate the perception, cognition, and communicative intent of a visualization \cite{agrawala2011design} by understanding the principles that make for a good design. Producing a ``friendly data graphic'' \cite{tufte1983visual} requires the designers to include a balance of useful features while also remaining aesthetically pleasing and not overwhelming to users. Well-chosen colors \cite{macdonald1999using}, fonts \cite{moere2011role}, and transitions \cite{heer2007animated} have been shown to affect usability of a visualization and have been strategically considered in the present work.

The combination of computer analytical processes with the human ability to quickly understand patterns leads to a reasoning system of discovery through interaction \cite{green_visual_2008}. This work focuses on compiling and arranging the tabular BLS data into an easily navigable interface that allows users to see trends in the data, from which to potentially draw immediate insights. The goal of these interactive visualization tools is to significantly improve the user interface of the current BLS website and benefit the entire labor statistics community.

\section{The BLS Data Pipeline}

The Jobs Report is released monthly based primarily on two national surveys: the Current Population Survey and the Current Employment Statistics Survey. Data was also ingested from state and metropolitan surveys: the Local Area Unemployment Statistics survey and Current Employment Statistics for State and Metropolitan economic divisions. Using these data sources created a unique challenge for a visual analytics data product - the system had to be able to adapt and acquire new information automatically, and to display important headlines on demand.

\begin{figure}[!h]
    \centering
    \includegraphics[width=0.9\textwidth]{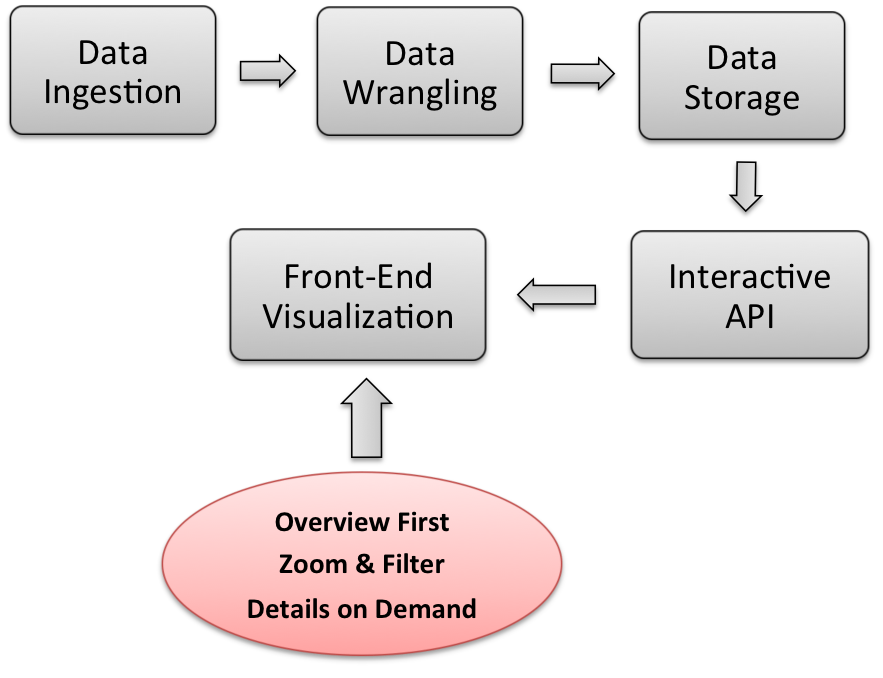}
    \caption{The BLS data pipeline, a model for combining monthly data ingestion with a visual analytics workflow.}~\label{fig:pipline}
\end{figure}

Our initial architecture therefore relied on the  ``data science pipeline'' \cite{ojeda_practical_2014}, an analytical framework, which includes preparatory analytical steps of ingestion, wrangling, storage, and pre-computation. Inferential and modeling statistical methods were replaced with visual analysis. The resulting ``BLS Data Pipeline'' is described by Figure~\ref{fig:pipline}. The ingestion component utilized the BLS API to download individual time series, which then had to be wrangled to normalize the data and input missing values for storage. Initial computations were then run on the data set to generate other data sources (e.g. percent change over time). The data storage then powered an interactive API that was utilized by our visual front-end.

\subsection{Data Sources}

The Bureau of Labor Statistics provides an expansive amount of data gathered by surveys conducted by various organizations of the United States Department of Labor. Several separate surveys, including the Current Population Survey, the Current Employment Statistics Survey, and the Local Area Unemployment Statistics Survey constitute the information that is used in our application to explore the employment situation of the United States at both the national and state levels. Each of these data sources adds dimensions to a variety of visual analyses, and when contextualized together provides the best possible opportunity for insight discovery.

The primary data source for the Jobs Report is the Current Population Survey (CPS) \cite{_labor_????}, which is a monthly survey of households. This survey focuses on the relationship between demographics and employment and, at the national scale, provides a comprehensive data set on the labor force, unemployment, earnings, hours of work, and other labor characteristics. This survey gathers demographic statistics like gender, age, education level, and ethnicity.

A secondary source for the Jobs report, the Current Employment Statistics Survey (CES) surveys employers rather than households for labor force information \cite{_current_2015}. The CES survey is not only national, but also provides granular data for state and metropolitan geographies as well. In the application presented here, national and local CES surveys are differentiated with the acronyms CES and CESSM, respectively. This survey focuses on industry-specific details of employment, hours, payroll, and earnings and is useful as a classification for different labor categories rather than demographics.

Finally, the Local Area Unemployment Statistics Survey (LAUS) fills in local regional information that relates demographic information to employment where the CPS only describes the national level \cite{_local_2015}. LAUS provides monthly employment, unemployment, and labor force data by census divisions that include states, counties, metropolitan areas, and many cities. Because LAUS is a household survey (by place of residence) the employment characteristics can give a much more focused view of how demographic trends influence employment at the local level.

Finally, datasets can be grouped by a variety of dimensions. The employment dimensions include ``employment'', ``unemployment'', ``unemployment rate'', and ``labor force''. CES dimensions included industry descriptions like ``Mining and Construction'', whereas CPS and LAUS focused on demographic dimensions like age, gender, ethnicity, or level of education. Geographic dimensions had to be included for the CESSM and LAUS data sets, ordered by both U.S. Census economic divisions as well as by state.

\subsection{Data Ingestion}

The Bureau of Labor statistics provides two public data APIs that use RESTful state transfer mechanisms to deliver data on demand to users for particular time series information in various BLS programs. The APIs accept HTTP requests for information to URL endpoints that describe the required resource, then return information in paginated JSON format. Version 2.0 implementation of the BLS API was used, which requires registration to obtain an API key. Such registration allows BLS to better understand what applications are using its data, but also allows developers more flexible access to the BLS data source.

A Python tool for downloading BLS data was created by providing a list of BLS IDs for each time series. There are three primary ways of ingesting time series: by downloading a single source or multiple sources to an endpoint that only allows downloading of data for the past three years, or by using the registration endpoint to specify the start and end years for ingesting multiple time series. Data was ingested for the current millennium: from January, 2000 to February, 2015, obtaining over 305,445 records from 1,684 time series. A custom web scraper obtained a list of time series from the "most popular" pages on the BLS website.

\subsection{Data Wrangling and Storage}

Although the data is returned in JSON format, each  data source from individual surveys may be formulated differently. Data wrangling efforts primarily dealt with inputing missing data (from years where surveys were not collected) and correcting mistakes in the ingestion process (e.g. mislabeled series information). Data type conversion was also required as JSON is a string format. The result of the wrangling process was a normalized, relational data set that could be stored in a PostgreSQL database management system.

The data wrangling system also computed ``extra'' data sources that might be interesting to the front end visualization. An example includes computing the percent rate of change of a surveyed category. We hoped that by including pre-computed data sets such as these, the visual exploration mechanism could combine both human and computer analytical efforts.

Data wrangling also involved initial data exploration efforts.  We discovered that in order to power the geographic analyses, the tool required some sort of regional information (e.g. FIPS identification codes or larger economic areas). The BLS website, while providing information particularly related to a time series, provided no \textit{meta information} about the data source, which required data retrieved from other sources. For example, although there might be a time series entitled ``Tennessee, Professional and Business Services, Not seasonally adjusted - labor force'' - this series had to be manually assigned the Tennessee FIPS code, 42, and parsed out whether or not the data set was seasonally adjusted, which dataset it related to (labor force), and which modifier should be used (Professional and Business Services). To do this for 1684 data sources, we built an automated mechanism to parse time series names. The challenges of adapting the data straight from the BLS website took up over 80\% of the development effort, and perhaps highlights how vital a visual analysis tool is for this type and quantity of data.

\subsection{System Architecture}

The ingestion mechanism can be polled on a monthly basis to retrieve new information from the BLS API, keeping the application up to date for immediate use by journalists. The data backend then powers a RESTful API built using Python and the Flask framework that serves a front-end javascript single page application, which primarily uses D3 for visualization methodologies. Other notable implementation details can be found on the Github project page: \url{https://github.com/bbengfort/jobs-report/}, where the authors have made our system open source using the MIT license. The project itself is currently deployed on Heroku and can be accessed via \url{http://elmr.herokuapp.com/}.

\section{First Inspirational Prototype}

\begin{figure*}[!ht]
    \centering
    \includegraphics[width=0.9\textwidth]{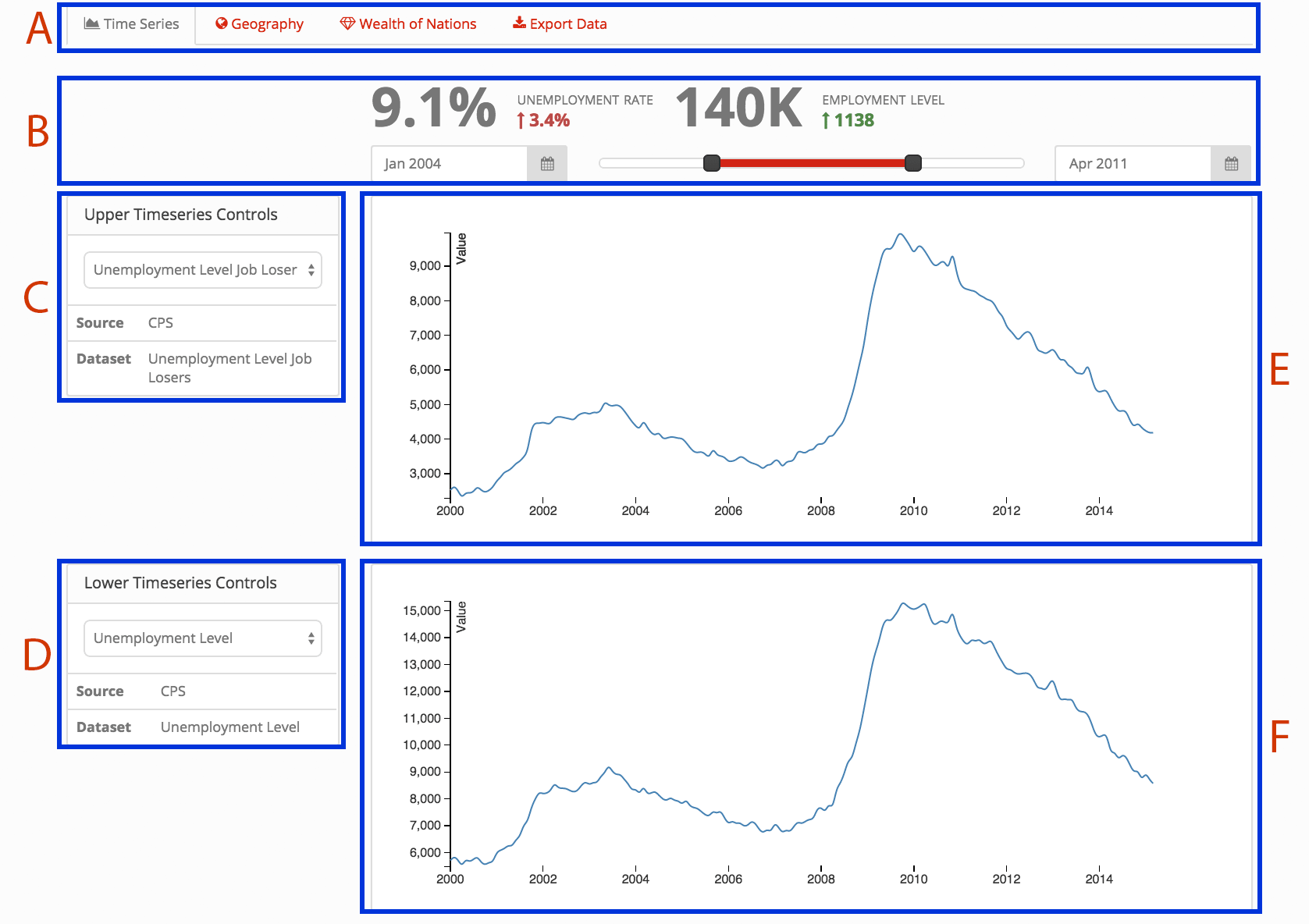}
    \caption{The time series explorer dashboard allows users to visualize multiple raw time series information in two different charts for ease of comparison with different data ranges. The year slider allows zooming and filtering by year. Finally, this dashboard attempts to add important headlines as listed by the Jobs Report.}
    \label{fig:time_series}
\end{figure*}

Our first inspirational prototype supports dynamic discovery in both time series and geography. Although each individual time series did not contain massive amounts of information (approximately 182 data points representing each month in this millenium), the number of time series themselves exceed the ability for a human to organize and digest the various information. In order to provide a visual mechanism of exploration and interactive reasoning, our design philosophy was that of a ``dashboard of visualizations'' where users could select appropriate visualizations, and interact with them though the ``overview first, zoom and filter, details on demand'' sense-making process. The primary filtering process involved time sliders to create windows of time to explore, as well as selectively choose which data to include in the visualizations.

Three dashboards were implemented to explore these techniques, each in their own separate tabs across the top of the application:

\begin{enumerate}
    \item \textbf{"Time Series" Explorer}: Explore and compare specific time series data sets in a dual-graph display.
    \item \textbf{"Geography" Choropleth}: Explore time series according to their regional factors and influences, and how regions change over time.
    \item \textbf{"Export Data" Utility}: Export data for use in another visualization or analysis tool. 
\end{enumerate}

The visual analysis process requires a ``dashboard'' methodology - where users can interactively control as many aspects of the visualization as possible, and where multiple visualizations are adapted simultaneously. Each tab described above is a dashboard with its own custom controls. The Time Series Explorer allows users to explore all of the available time series with two chart windows and filter and zoom by a range of years. It also provides a ``headline view'' which attempts to highlight key information, such as percent change over a selected time period. The Geography Choropleth allows users to explore regional information in the continental United States, while also filtering for percent change. Finally, a data export dashboard allows users to download specific data with filters applied so that they can manipulate the data in their favorite visualization tools.

\subsection{Time Series Explorer}

Figure~\ref{fig:time_series} provides a snapshot of our design for the time series dashboard. The primary focus of this dashboard is two vertically stacked time series graphs, which allow for a visualization of multiple datasets at once. This is especially helpful if the two quantities are of different dimensional units (e.g. population and unemployment rate) and could not otherwise be plotted together. In order to better describe the dashboard, it was divided into six regions labeled A-F.

\begin{figure*}[ht]
    \centering
    \includegraphics[width=0.9\textwidth]{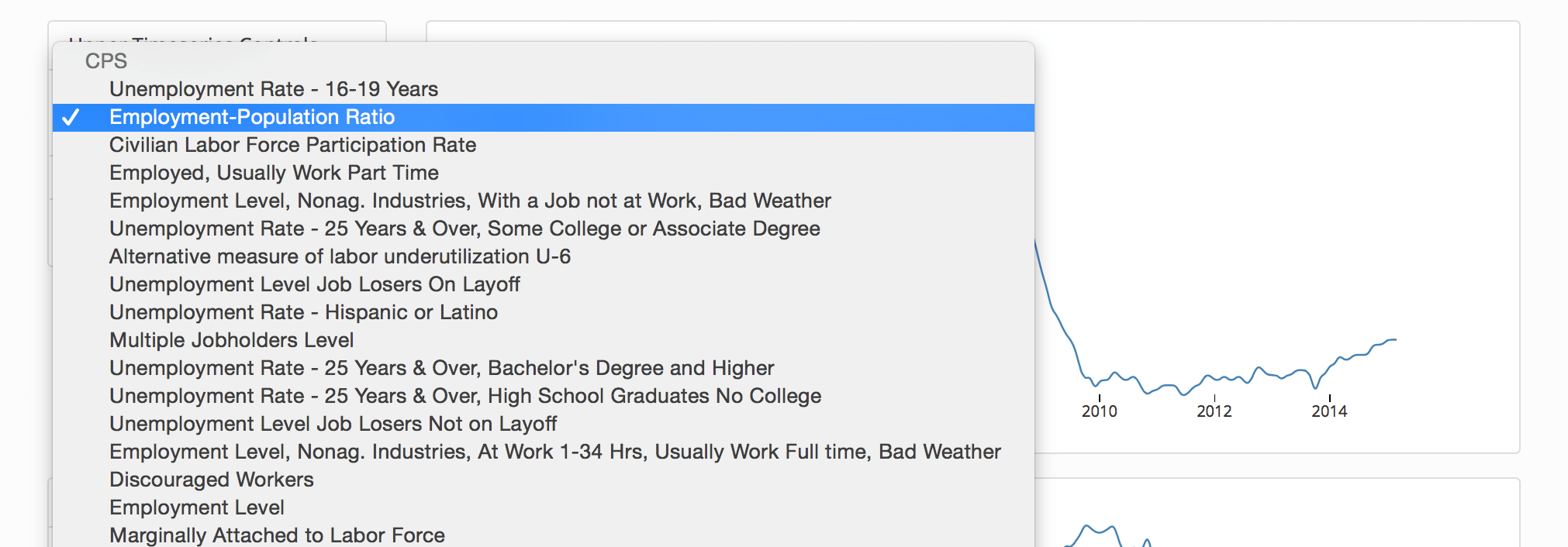}
    \caption{In order to minimize error, users select bls time series data sets through a grouped, alphabetized dropdown.}
    \label{fig:select}
\end{figure*}

Region A shows tab navigation bars which allow users to switch between the three dashboards. The tab display is consistent across all dashboards, and as more visualizations are added, more tabs can be added as well. If the list of tabs grows too long, the authors intend to allow users to select tabs for a personalized display that gives them quick access to the dashboards they use most (for example, national reporters may be less interested in local employment statistics).

Region B contains an ``at-a-glance headline view'' and year slider control. The headline view is intended to be an eye-catching ``at-a-glance'' instant insight framework that is based heavily on the headline in the Employment Situation Report. The headline is tied to the two major pieces of information that are published at the top of the report - the unemployment rate and the number of non-farm jobs. The at-a-glance headline also displays important information regarding the amount of \textit{change} from the last period - a major piece of information that most users want to know. It is for this reason that these headlines have such a prominent place in our application.

The year slider control also shown in Region B is the primary mechanism for zooming and filtering information for every dashboard in the app. The year range slider allows users to select a time period within the available time span and zoom the data into that range. Other dashboards have period selectors rather than ranges, which allows users to identify trends or data for a specific period.

Regions C and D contain the primary controls for the time series displays in regions E and F, respectively. This application contains 1684 downloaded data sets, each of which can be displayed. The challenge is how to allow users to select data sets of interest quickly. Although using an autocomplete or search mechanism was considered, the prototype implementation simply uses a grouped, auto complete dropdown as shown in Figure~\ref{fig:select}. This design follows ``prevent error'' rule in the Eight Golden Rules of User Interface Design \cite{shneiderman_eight_????} as users can only select the databases that are available through our drop-down menu.

\subsection{Geography Choropleth}

\begin{figure*}[!ht]
    \centering
    \includegraphics[width=0.9\textwidth]{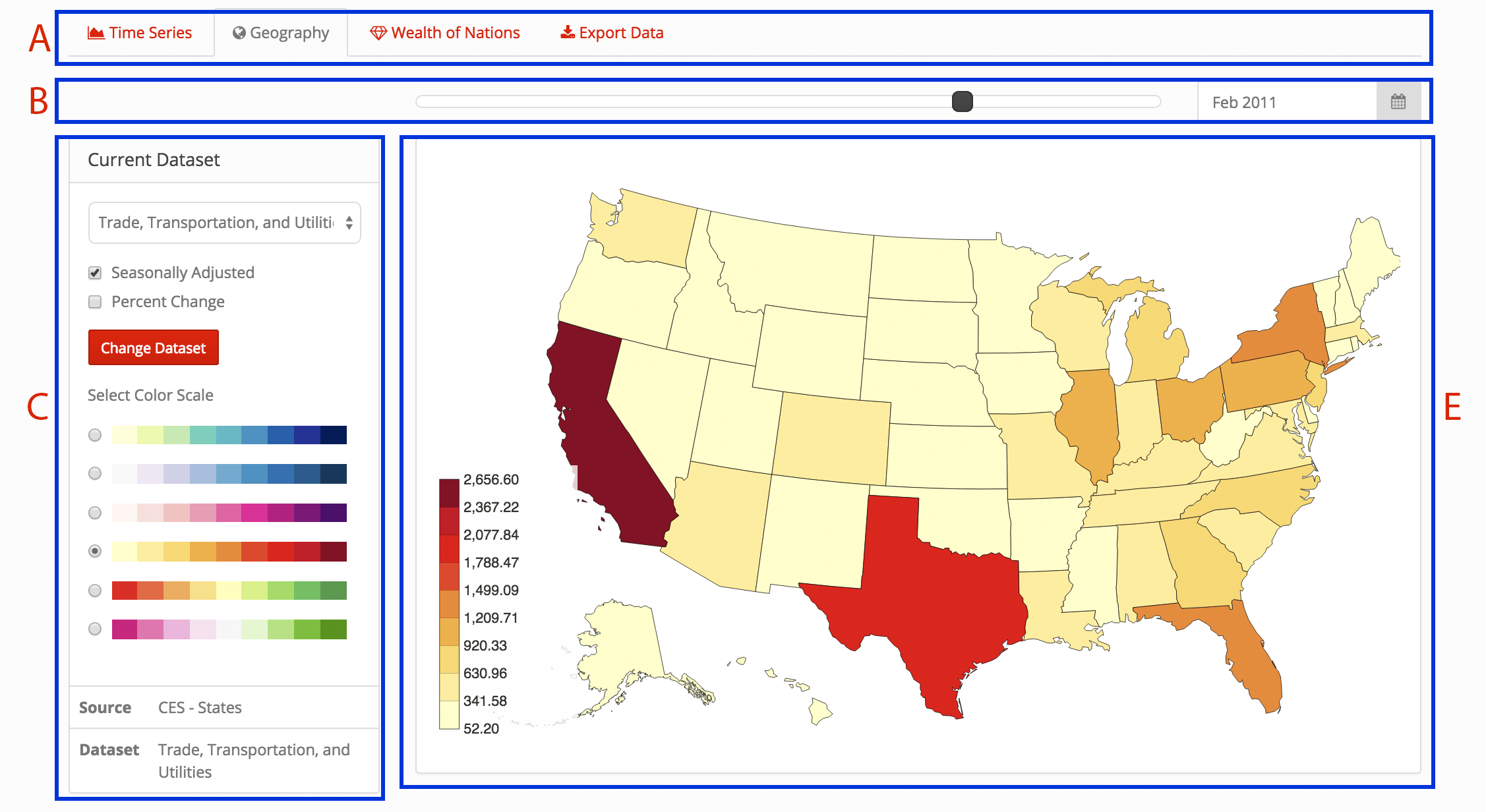}
    \caption{The Geography Choropleth dashboard implements a chorpleth style heatmap on the AlbersUSA projection of the United States, particularly at the state and regional level. Users can control the data set and colors with the control bar, and select the time period to visualize with a slider.}
    \label{fig:Geography}
\end{figure*}

Our second visual dashboard focused particularly on regional and state visual analyses. It implements a highly interactive choropleth map of the United States using the AlbersUSA projection, as shown in Figure~\ref{fig:Geography}. The dominant feature of this dashboard is the choropleth map of the United States which is directly connected to a selectable dataset from either the CESSM or LAUS datasets. By interacting with the controls on the left, users can select the data set, whether or not to implement seasonal adjustments, choose to show the rate of change calculation, and even select the color map. A slider at the top of the visualization controls the year for the map.

This style map was the first design for the application, and its implementation was large focus of the efforts. However, we observed that the datasets were heavily weighted towards the larger states of California and Texas, reducing the usability of a choropleth, which requires a color scale to be effective. Data was normalized using "percent change" computations as described in the wrangling section. In this case users can explore the magnitude of change in a time series per state, which is much more effective in the choropleth setting.

Region C is designed for users to select and adapt the data set they want to visualize. A grouped, alphabetized drop-down menu is used for users to select a data source from either LAUS or CESSM. Two check boxes allow users to specify whether they want to show ``seasonally adjusted'' data and/or the percentage of change from the previous month. Then users click on the ``Change Database'' button to refresh the map. To offer users flexibility, the application provides  color scales to plot the data. Depending on the specific quantity plotted,  users may prefer a particular form of color scale, e.g. sequential, diverging, etc.

Region E is the main choropleth that displays the continental United States as well as Alaska and Hawaii. The legend shows the color scale at the bottom left. If users hover over an individual state, a tooltip pops up with detailed information. Users can also click on a state to zoom in and see specific information about that state.

\subsection{Data Export}

Finally, the last dashboard implements a data exporting feature so users can use other software, such as Excel or Tableau, to perform their analyses. The ingestion and wrangling of time series data is not difficult, but it is time consuming. By allowing users to directly export data from the tool, we hope to increase the number of analyses available and the insights that can be gained. Datasets are then downloaded in CSV or JSON format.

\subsection{Data Management}

\begin{figure*}[h]
    \centering
    \includegraphics[width=0.9\textwidth]{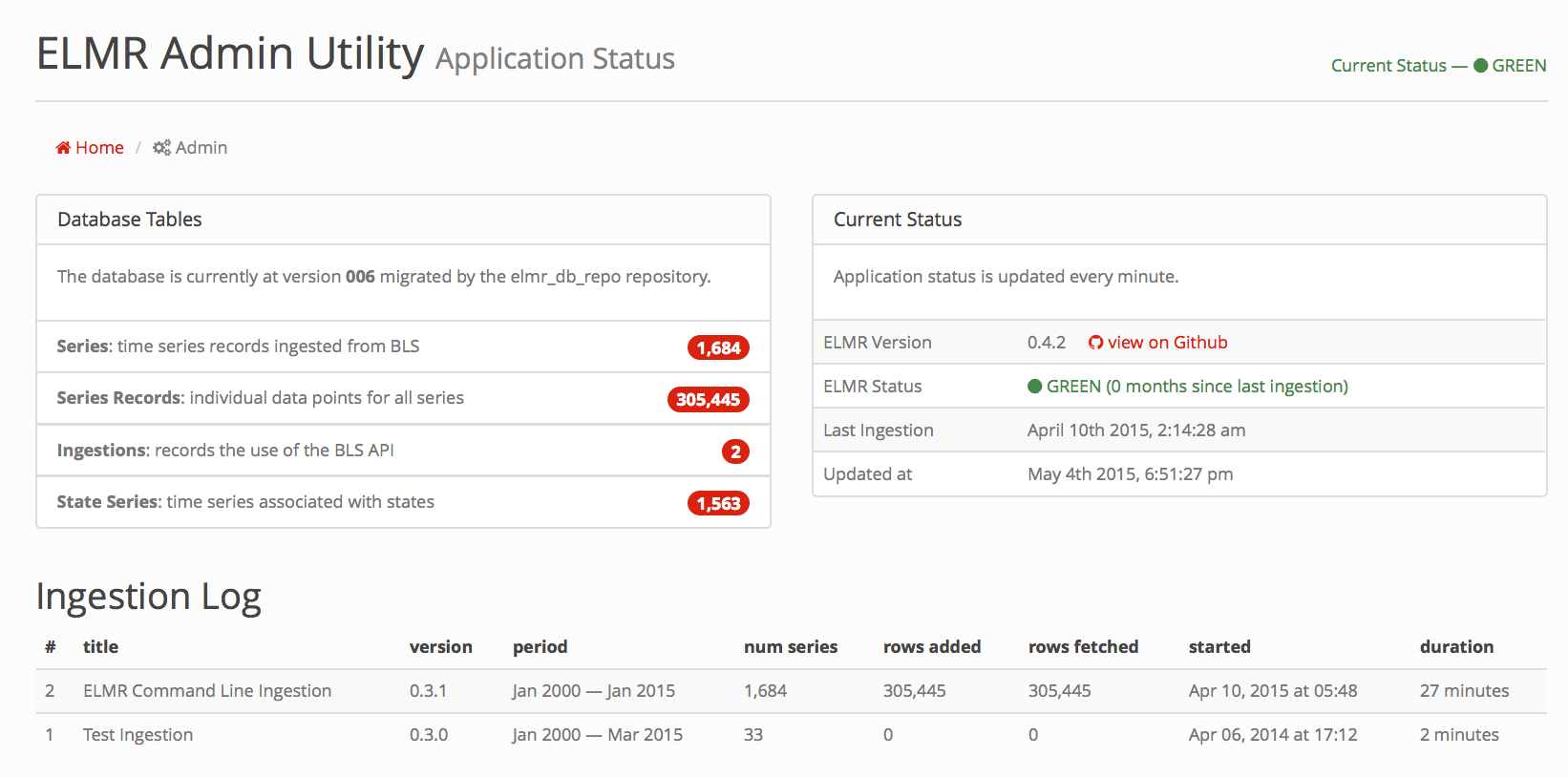}
    \caption{Illustration of Admin resources.}
    \label{fig:admin}
\end{figure*}

Because the ingestion utility is real-time and visual analysis dashboards will change over time, an admin screen gives users an at-a-glance look at the system state (Figure~\ref{fig:admin}). This page conveys the effort it took to wrangle and gather the data and provides a means for interacting with the real-time ingestion system. For example, an ingestion log shows when data was fetched from the BLS web site, as well as the magnitude of the ingestion process.

Other details include the current version of the application and the timestamps of the updates or ingestions. The status bar gives a quick view of any required maintenance for the system. A GREEN status indicates that the app is within a month of the last ingestion, YELLOW indicates that the app is less than six months since ingestion, and RED indicates that the app has not been updated in over six months. Database records also indicate the state and size of the database.

Creating an open utility that shows the ingestion process and the database state allows users to view the analytic process, encoraging collaborative efforts to improve the system. Our software is open source on Github, with links to full documentation, agile development boards, API documentation, and more. Future goals are to allow direct administration from this page, by kicking off ingestion, or editing time series directly.

\section{Second Inspirational Prototype}

The dashboard applications in the previous sections provide a foundation from which to assess the strengths and weaknesses of visualizing the employment report. Through iterative testing and redesign, we developed a second inspirational prototype for organizing and plotting the time series data.  A new tool, titled BLSVisualizer, allows users to clearly and interactively navigate through the employment data, which is conveniently organized in a hierarchical format. 

The main feature of BLSVisualizer is the use of an interactive tree structure to locate data. The tree structure is generated by implementing a parent-children hierarchy, from which users can select their her desired dataset. Much work has gone into studying the effectiveness of tree structures from both practical \cite{barlow2001comparison, plaisant2002spacetree} and aesthetic \cite{wetherell1979tidy, reingold1981tidier} standpoints. Barlow and Neville performed a parameter study on hierarchy visualizations by asking users to perform decision tree analyses on four different chart types: Organizational chart (or tree structure), icicle plot, tree ring, and treemap. Based on user response time and response accuracy, the authors concluded that the tree structure and icicle plots produced the most effective visualizations. Therefore, we implemented the hierarchical tree structure, due to its widespread familiarity, ease of implementation, and aesthetic appeal.

The second feature of BLSVisualizer is the real-time drag-and-drop plot area, where users can view the time series data in the form of two-dimensional line charts. While previously users were forced to deal with hundreds of tables of data to compile a single time series, users can now locate and plot the data quickly, ideally leading to more efficient data analysis.

Data was acquired via the API provided by the BLS in comma separated value (CSV) format. The CSV formatted data was converted to JSON format in order to more easily implement the hierarchical parent-child tree structure.

\begin{figure*}[t]
\centering
    \includegraphics[width=1\textwidth]{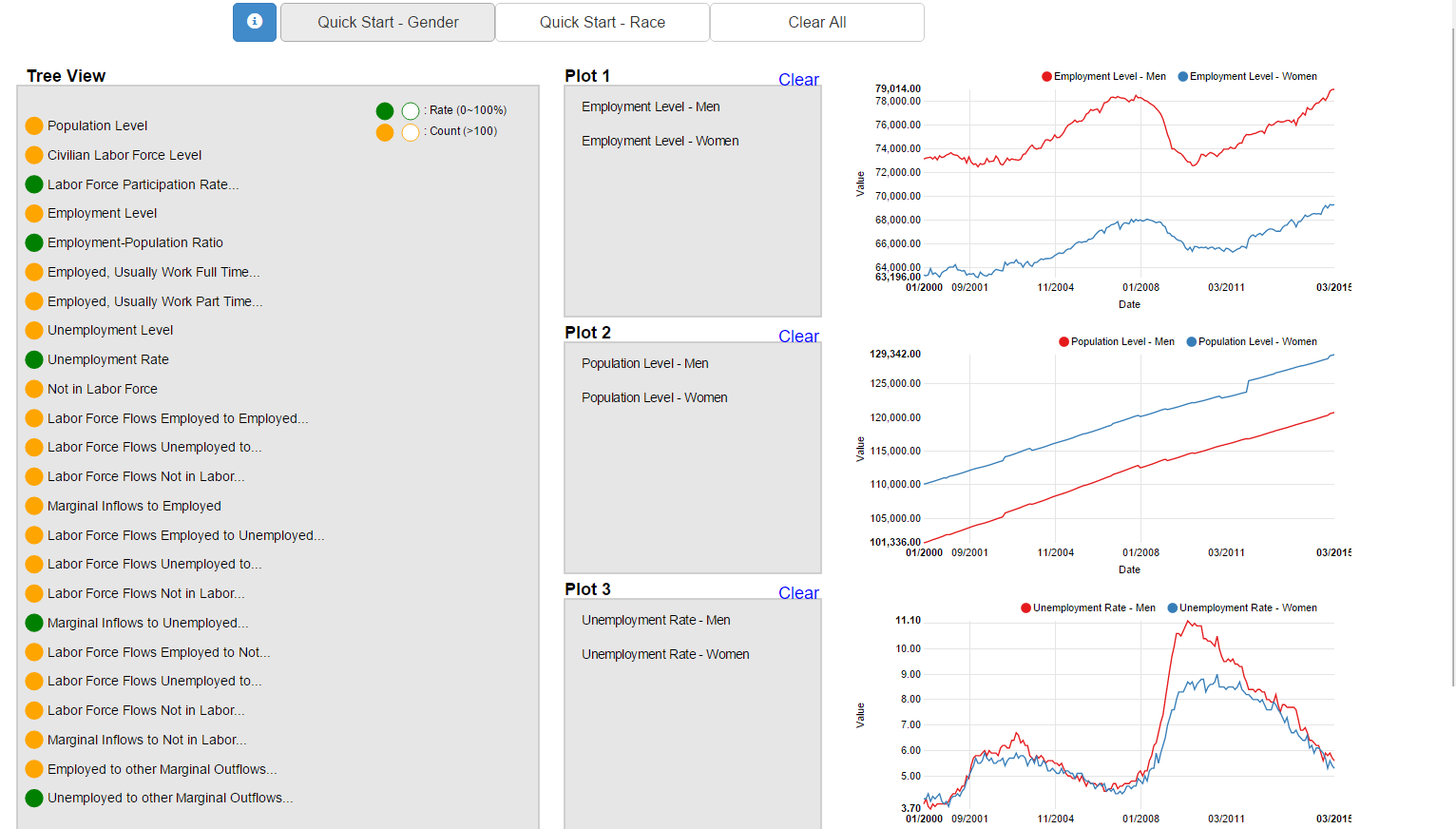}
    \caption{Example layout of BLSVisualizer }
    \label{BLSVLayout}
\end{figure*}


\subsection{Tree Structure}

A primary step in any design process is to identify the design space. In this case, the space consists of multi-leveled categorical data. The nature of the design space lends itself to be organized in a hierarchical format. Hierarchies are typically displayed using a geometrically-transformed display \cite{keim2002information}. This is done in the present work in the form of an interactively transforming tree structure, primarily written using the D3 library titled ``Collapsible Tree Layout''. The smooth transitions are meant to be aesthetically pleasing as well as visually effective.

Wetherell and Shannon \cite{wetherell1979tidy} provide a detailed description on what makes for an effective tree visualization. They identify three key \textit{aesthetics}, which were implemented into BLSVisualizer:

\begin{itemize}
\item \textit{Aesthetic 1}: Nodes of a tree at the same level should be along a straight line
\item \textit{Aesthetic 2}: Left and right children should remain left and right of the parent
\item \textit{Aesthetic 3}: Parent should be centered over its children
\end{itemize}

Another feature Wetherell and Shannon suggested was to minimize the width of the tree as much as possible. However, the present authors and Reingold et al. \cite{reingold1981tidier} found this technique to cause a cluttered workspace and too narrow of a visualization.

Most of the data, when exported from the BLS website, comes organized in a conveniently hierarchical manner. However, data was further condensed to be organized in a more logical manner under more appropriate parent levels. To maintain a clutter-free environment, as users expand the tree, only the bottom-most levels are visible. To return to a previous level, users can re-click on the last expanded parent, collapsing the bottom level and returning the previous level into view. The colors of each item correspond to either dimensional values (orange) or non-dimensional values (green), such as rates or percentages. Since this is a rather large dataset, this minor color coding will help users more quickly identify the dataset they desire. The entire tree interface specifically complies with three of the Eight Golden Rules \cite{shneiderman_eight_????}: Strive for consistency, easy reversal of actions, and reduce short-term memory load.

\begin{figure}[h]
\centering
    \includegraphics[width=1\textwidth]{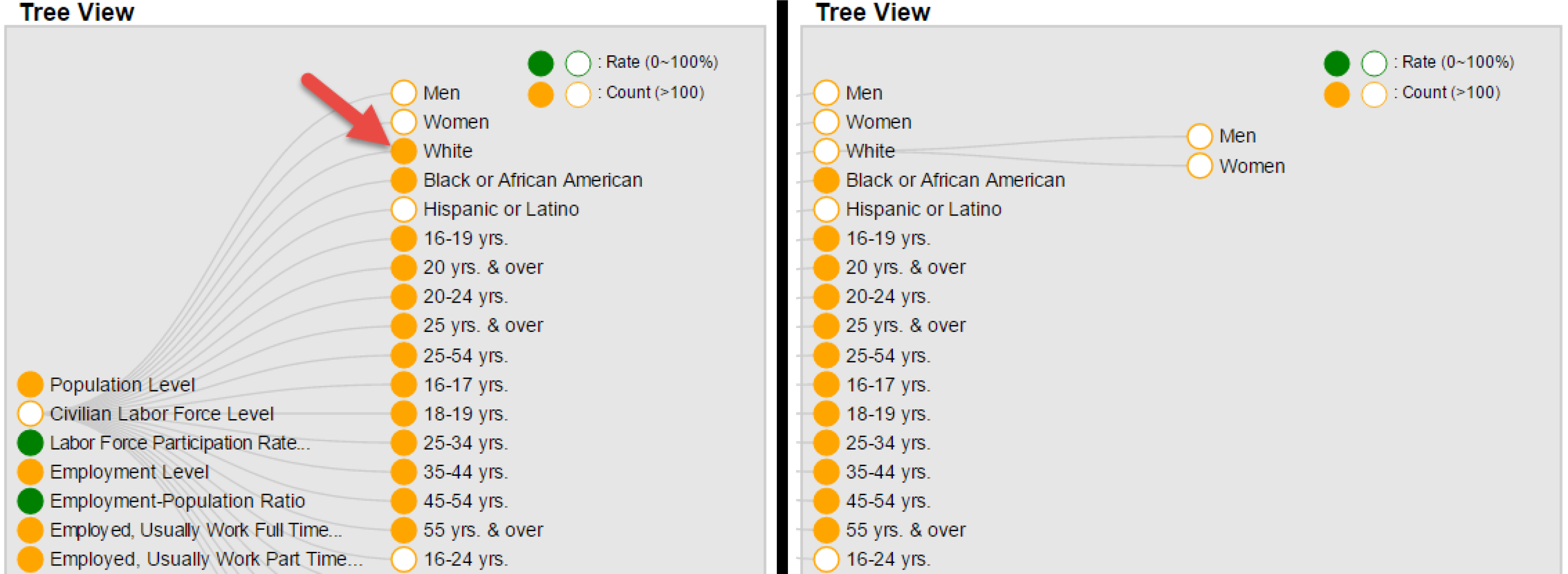}
    \caption{Interactive hierarchical tree structure. (Left) One level into the hierarchy and (right) two levels into the hierarchy, during which the first level is pushed off-screen.}
    \label{TreeStructure}
\end{figure}

\subsection{Drag-and-Drop Plot}

To plot the time series data, users select a particular category and drag-and-drop that node into one of the middle boxes labeled ``Plot 1-3''. Each plot allows up to six items to be plotted at any one time. In order to preserve an easy reversal of actions, users may simply click on the item in the plotting list to remove it from the chart. The ``clear'' button at the top of each plotting area provides a one-click option to clear the plot of all items. Again in compliance with the Eight Golden Rules, in order to ``prevent errors'' the application prevents users from adding the same category more than once to the same chart.

\subsection{Chart Area}

The most important component of the visualization is the time series chart area. Dragging components into the center boxes causes the charts on the right to interactively update with the currently listed items. The time series data suggests that the shape of the graphic be a rectangular plot with the horizontal axis containing the date of each released BLS report. Line color was selected using the cartography tool ColorBrewer \cite{harrower2003colorbrewer}. This tool provided an array of six colors, arranged with the intention of plotting data with qualitative differences. This is opposed to a quantitative color scale, e.g. light red to dark red, that would indicate the quantitative relationship between the items plotted. However, the two datasets of ``Employment Level - Men'' and ``Employment Level - Women'', for example, have no quantitative relationship. Thus, a color scale that provides significant contrast and legibility between the individual lines is ideal.

BLSVisualizer, as a whole, complies well with the popular mantra for designing an effective visualization of ``Overview first, zoom and filter, then details-on-demand'' \cite{shneiderman_eyes_1996}. The tree itself acts as an overview-to-zoom interactive element, allowing users to dive deeper into the data, getting more detailed with each subsequent level. Additionally, the line charts, plotted from years 2000 to 2015 provide an overview in the sense that it is a line showing the trend of data over 15 years. However, upon hovering over the plot, a tooltip box provides the discrete monthly data points at the current cursor location. To help users identify where along the x-axis the cursor is located, a vertical line spanning the height of the plot follows the cursor as it travels horizontally across the figure. The vertical position-indicator line is colored gray and is significantly thinner than the plotted time series data. The contrast in line weight represents a contrast in meaning, with the greater meaning given to the line with greater weight \cite{tufte1983visual}.

\begin{figure*}[h]
\centering
    \includegraphics[width=1\textwidth]{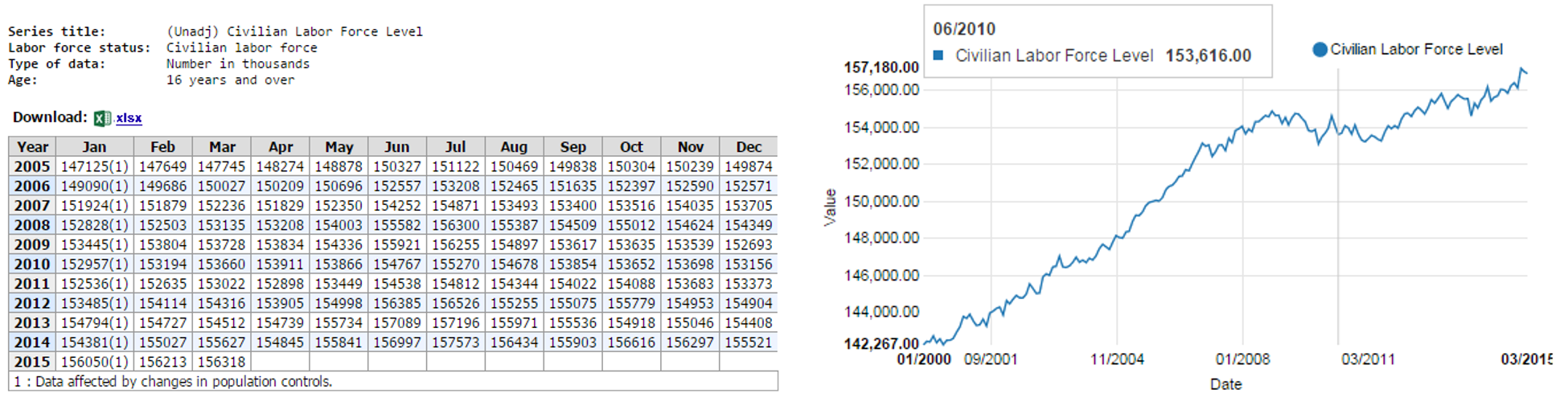}
    \caption{Comparison of the two proposed sets of BLS data. (Left) Tabular data presently employed on the BLS website and (right) its equivalent visual counterpart presented in BLSVisualizer.}
    \label{TablePlot}
\end{figure*}

Figure \ref{TablePlot} provides a comparison of how the data is presented in both the current BLS website and in the BLSVisualizer. It is clearly shown that BLSVisualizer provides a service in terms of observational insightfulness when viewing the data that the current BLS website does not provide.

\subsection{Usability Study}

To examine BLSVisualizer, remote and in-person usability tests were performed with five users, about whom we collected the basic information of age, gender, occupation, and experience with visualization tools. All five were students, but only three participants had previous experience with visualization tools. Due to the relative simplicity of the tool's user interface, which was the primary focus of the study, a small sample size of five users with varying technological backgrounds was deemed sufficient. 

The 5-10 minute test was designed to progressively increase in difficulty, and went as follows:

\begin{enumerate}
  \item Create 3 graphs for the following categories:

\begin{itemize}
  \item Population level
  \item Unemployment level
  \item Marginal inflows to employment
\end{itemize}

  \item Change ``Unemployment level'' to ``Employment level''
  \item What was the white population level at 11/2011?
  \item Between 1/2008 to 1/2010, who had a larger increase in labor force flows from unemployment to employment: Men or Women?
  \item Compare, on the same graph, the effect of education level on unemployment rate.
  \item Compare the employment level of four different industries (both sexes).
\begin{itemize}
  \item Identify the largest industry (in terms of employment level) and compare the distribution of men and women for that industry.
\end{itemize}

\end{enumerate}

Question \#1 simply asks users to plot three separate datasets, which are all located on the first level of the tree, on the three chart areas. After reading the ``About'' section at the top of the page that describes how to use the tool, none of the users had an issue plotting the data. Two users brought up the fact that they were not aware that the ``About'' button was even an option. A clearer button label would have remedied that issue.

Question \#2 asks users to remove one of the datasets they plotted and replace it with another. This question tests the ease of undoing an action, which is a very important feature in a visualization tool. Again, after locating the ``About'' pop-up window, all users were able to easily perform this task.

Question \#3 tests the plausibility of replacing the current table look-up style of presenting the data with this interactive visualization. This task asked users to plot a time series and identify its value at a particular date (month/year). This task can be done in BLSVisualizer in two clicks, whereas on the BLS website it would likely take up to six clicks while visually sifting through dataset titles.

Question \#4 attempts to do what the current BLS data presentation cannot provide: Insights based on data visualization. Users are asked to identify a way to compare the effect of education level on unemployment rate. This question gave several users some trouble. They were not sure what exactly they were asked to do, but after exploring the categories and understanding what options they had to work with, all five users eventually produced the correct plot and drew the same conclusion. The conclusion was that the more educated you are the less likely you are to be unemployed.

The final task of the usability test asked users to plot the employment level of four industries of their choice and identify the one with the highest level of employment. Once this dataset was located, they were asked to break that data down into a demographic study of male and female. This question was created to show users that this visualization promotes breaking the data down into its parts to identify sources of a particular trend in the data. Users had minor difficulties with this task; however, once the task was completed, their comments conveyed that they were impressed with the overall utility of the tool. One of the test users, an Australian exchange journalism student, commented: "One major aspect of journalism is detecting changes in trends over time, or noticing irregular disparities between two sub-groups of society. This can be quite difficult to do, due to either a lack or disorganization of relevant data. This [visualization tool] allows you to quickly access data, collect it into one place, and easily divide it into subgroups to visualize trends. Journalists can use this to pick up on interesting irregularities in the data and create a story about it." 

The most common criticisms and suggestions to improve the BLSVisualizer interface were:

\begin{itemize}
  \item Search bar
  \item Option to add more plots
  \item Option to export data
  \item Increased font size
  \item Category titles that do not overlap other text
\end{itemize}

\section{Conclusions and Future Work}

Our work focuses on improving the user interface of the current Bureau of Labor Statistics website by compiling and arranging the fixed tabular BLS data into an easily navigable interface to support dynamic discovery of trends in the data. The tool aggregates a large dataset of over a thousand time series, ingested using the BLS API on a monthly basis. 

We implemented two inspirational prototypes: a Geographic Choropleth map to investigate changes in employment according to demographics or industry at the state level, and a Time Series explorer, where users can compare and contrast multiple time series through a stacked overlay. These dashboard applications, made primarily through the use of the D3 Javascript library,  provided a foundation from which to assess the viability of an interactive visualization for this dataset. Through iterative designs and usability tests, we determined that visualization tools improved the user experience. Taking the initial dashboard tool a step further and improving upon the time series application, a hierarchical tree structure was introduced for more intuitive navigation. Additionally, a drag-and-drop method of plotting was introduced to give users control over where the data was plotted and in what order. The chart area was also refined to provide discrete values at the point where the cursor is located. 

A small usability test with five users concluded that this application is more useful, in terms of both finding the desired data and gathering conclusions based on trends. Usability testing also highlighted several areas of improvement for the visualization. The font and inter-category spacing will be made larger for improved legibility. Also, the ``Help/About'' sections will be more clear as to what capabilities the application has to offer. The most prominent user feedback, however, was the desire for a search bar to locate particular datasets.

This work provides a step toward making the BLS website more accessible to a wider range of users. The inspirational prototypes presented here provide a base design  to build upon, and the feedback and observations  should be considered when building dynamic discovery tools to supplement the current BLS website.

\section{Acknowledgments}

This work was a team effort and many people collaborated on both the ELMR and BLSVisualizer projects. We would especially like to thank Bor-Chun Chen, Xintong Han, Jonggi Hong, Rotem Katzir, Assaf Magen, and Hao Zhou for all their support. Together they assisted us with software development, usability studies, and creating videos and demos. We would also like to thank those who participated in our usability study, and who critiqued our work.

We would also like to thank Jonathan Schwabish of the Urban Institute who provided the framework for the project and the Bureau of Labor Statistics staff, particularly Emily Liddel and Tyrone Grandison who gave us guidance with the BLS API, offered specific information about the data sets being used, and arranged for our presentation at BLS to Commissioner Groshen and staff. 

\section{Links and Resources}

The following links contain more information relevant to the project as well as resources for further exploration (they have been shortened by the ter.ps URL shortner project for print format).

\begin{enumerate}
    \item ELMR Application: \url{http://ter.ps/985}
    \item ELMR Video Tutorial: \url{http://ter.ps/983}
    \item BLSVisualizer Application: \url{http://ter.ps/9zu}
    \item BLSVisualizer Tutorial: \url{http://ter.ps/9zv}
    \item Github Repository: \url{http://ter.ps/984}
    \item Project Wiki: \url{http://ter.ps/986}
\end{enumerate}

\bibliographystyle{apacite}
\bibliography{paper}

\end{document}